\documentclass[10pt,letterpaper]{article}
\usepackage{opex3}
\usepackage{cite}
\usepackage{graphics}

\usepackage{amsfonts,amsmath,graphicx,color}
\usepackage[english]{babel}

\begin{document}
\title{Doubly resonant optical nanoantenna arrays for polarization resolved measurements of surface-enhanced Raman scattering}

\date{\today}
\author{J. Petschulat$^{1\dagger}$, D. Cialla$^{2,3}$, N. Janunts$^{1}$, C. Rockstuhl$^{4}$, U. H\"ubner$^{2}$,\\
R. M\"oller$^{2,3}$, H. Schneidewind$^{2}$, R. Mattheis$^{2}$,\\
 J. Popp$^{2,3}$ A. T\"unnermann$^{1}$, F. Lederer$^{4}$, and T. Pertsch$^{1}$}
\address{
$^\textbf{1}$ Institute of Applied Physics, Nanooptics, Friedrich-Schiller-Universit\"at Jena, Max Wien Platz 1, 07743 Jena, Germany \\
$^\textbf{2}$ Institute of Photonic Technology, Albert Einstein Stra\ss e 9, 07745 Jena, Germany \\
$^\textbf{3}$ Institute for Physical Chemistry, Friedrich-Schiller-Universit\"at Jena, Helmholtzweg 4, 07743 Jena, Germany, \\
$^\textbf{4}$ Institute of Condensed Matter Theory and Solid State Optics, Friedrich-Schiller-Universit\"at Jena, Max Wien Platz 1, 07743 Jena, Germany}

\email{$\dagger$ joerg.petschulat@uni-jena.de}

\begin{abstract} We report that rhomb-shaped metal nanoantenna arrays support
multiple plasmonic resonances, making them favorable bio-sensing
substrates. Besides the two localized plasmonic dipole modes
associated with the two principle axes of the rhombi, the sample
supports an additional grating-induced surface plasmon polariton
resonance. The plasmonic properties of all modes are carefully
studied by far-field measurements together with numerical and analytical
calculations. The sample is then applied to surface-enhanced Raman scattering
measurements. It is shown to be highly efficient since two plasmonic
resonances of the structure were simultaneously tuned to coincide with the
excitation and the emission wavelength in the SERS experiment. The analysis is completed by measuring the impact of
the polarization angle on the SERS signal. 
\end{abstract}

\ocis{(250.5403) Plasmonics, (240.6680) Surface plasmons, (160.3918) Metamaterials, (240.6695) Surface-enhanced Raman scattering}

\section{Introduction}
The use of metal nanoantennas is superior if one aims at localizing
light in volumes much smaller than the diffraction limit. The
physical reason for this behavior is the excitation of either
localized surface plasmon polaritons (LSPP) or propagating surface
plasmon polaritons (PSPP) \cite{Kreibig,Raether}. Both are
collective oscillations of the charge density at the interface
between a metal and a dielectric that is driven into resonance by an
illuminating wave field. It causes strongly localized
electromagnetic fields that decay evanescently off the
interface. PSPPs are mainly confined along one \cite{Barnes,Tejeira}
or two dimensions \cite{Bozhevolnyi1,Bozhevolniy2}. LSPPs are
confined in all three spatial dimensions. It has been shown that for
LSPPs excited at nanoantennas comprising sharp edges or coupled
systems thereof
\cite{Nanoprisms,Sanchez,vanDuyne,Muehlschlegel,Nordlander1,Huang}
as well as for PSPPs at interfaces the field localization is
accompanied with a dramatic field enhancement \cite{Russell}. This
enhancement can be used, e.g., to boost the efficiency of nonlinear
optical frequency conversion \cite{NonlinNL1} or of
surface-enhanced Raman scattering (SERS) applications
\cite{Dana,Hering,Dana2,SERSNL1,Kneipp,SERSangular}. Consequently the combination of both effects (LSPP and PSPP resonances) has been shown to improve the capability of such devices furthermore \cite{Baumberg1,Baumberg2,Baumberg3,Baumberg4,Baumberg5,empty_lattice}. Therein metal surfaces incorporating hexagonal arranged metallized nanovoids have been shown experimentally and theoretically to exhibit both LSPP (void plasmons) and PSPP modes.\\
In our contribution we present a combination of LSPP and PSPP excitations for two-dimensional sharp-edged rhombic nanoantenna arrays.
A LSPPs can be excited as fundamental plasmonic dipole modes of the individual nanoantennas. A PSPP can be excited because of the periodic arrangement of the nanoantennas. The most relevant difference of such structures to the previously mentioned approach is the absence of a closed metal surface in our work. Compared to the LSPP and PSPP combinations for structures incorporating such a closed metal surface \cite{Baumberg1,Baumberg2,Baumberg3,Baumberg4,Baumberg5,empty_lattice} our samples can be understood as a corresponding inverse approach. Thus the quasi-PSPP mode reported here propagates at the interface of an artificial effective medium whose dispersive properties are dictated by the isolated nanoantenna. This approach permits an individual tuning of two plasmonic resonances to a specified spectral regime. The LSPP mode can be controlled by tailoring the particle's shape and choosing the dispersive material properties it is made of; while the PSPP resonance is predominantly adjusted by selecting a certain grating period. \\
An alternative approach for substrates accessing multiple plasmon resonances consists of using nanoantennas designed to support next to a dipolar resonance also higher order plasmonic eigenmodes (e.g. the electric quadrupole mode). For example Ref.~\cite{vanDuyne,referee2_2} and \cite{referee2_3} also aim at realizing more than one resonant plasmonic resonance within the same nanostructure. By virtue of their multiple plasmon resonances these structures also provide  a dramatic field enhancement making them promising SERS substrates. The difference to the approach presented here is that the resonance positions of these localized modes (higher order localized modes as well as LSPPs and PSPPs in nanosphere lithography manufactured samples) cannot be chosen independently, since they are both controlled by the individual nanoparticle geometry. In contrast the samples introduced herein allow for an individual adjustment of the spectral properties of both the localized and propagating plasmon. Furthermore, our structures are fabricated by electron beam lithography that allows  to fabricate large areas of samples with reproducible properties suitable for the perspective of more industrial oriented SERS applications.
This
combination of LSPPs and PSPPs in a definite selected spectral interval
results in an improvement of near field sensitive plasmonic effects,
e.g. the SERS enhancement. We present a detailed numerical
investigation of the optical properties of our sample that is in
excellent agreement with the experimental spectral analysis. In
order to separate the effects of the appearing LSPPs from PSPPs, we
present also a detailed analytical electrodynamic model applying
well-known classical descriptions of plasmonic nanostructures. The
combination of the numerical and the analytical description
facilitates the understanding of the plasmonic origin of the
respective resonances and
allows for a simplified, yet precise, prediction of the optical response of such samples.\\
Their unique feature to significantly enhance the
efficiency of SERS measurements is furthermore experimentally proven.
This is possible since the resonances were adjusted to coincide
simultaneously with the excitation and the emission wavelength. Thereby we observe that the quasi-PSPP mode yields a similar SERS enhancement as it is achieved by localized modes making the combination of both a promising approach for improved SERS substrates. For
providing a comprehensive understanding of our sample we put also
emphasis on detailing the angle dependent SERS signals. It is used
to elucidate the interplay between the plasmonic
properties and the SERS signal.\\
\section{Fabrication}
It has been shown in earlier publications that random-rough metal surfaces exhibit a
field enhancement due to their sharp-edges
\cite{Albrecht,Jeanmarie,Fleischmann}. This field enhancement reported in these pioneering SERS papers is now understood to be
the origin of the enhancement of the Raman signals if molecules are
brought in close proximity to such surfaces; coining the term
surface-enhanced Raman scattering. Since then various technical approaches have been realized to achieve dramatic near field enhancements accessing even the single molecule detection. A simple strategy to manufacture large area samples supporting such features is to apply self-organization processes.
However, the intrinsic drawback
of such random-rough surfaces is their non-deterministic character.
Thereby it is difficult to observe a reliable and reproducible
enhancement of the SERS signal over large surfaces for such substrates. Moreover, a
large fraction of the surface does not contribute to the SERS signal
as it acts non-resonant. Thus, we are interested in realizing
artificial and deterministic SERS substrates comprising nanoantennas
with sharp edges that sustain resonances in well-defined spectral
domains. Ideally, the fabrication shall be simple and reliable to
allow for an up-scaling of the fabrication
process.\\
\begin{figure}
\centering
\includegraphics[height=3cm]{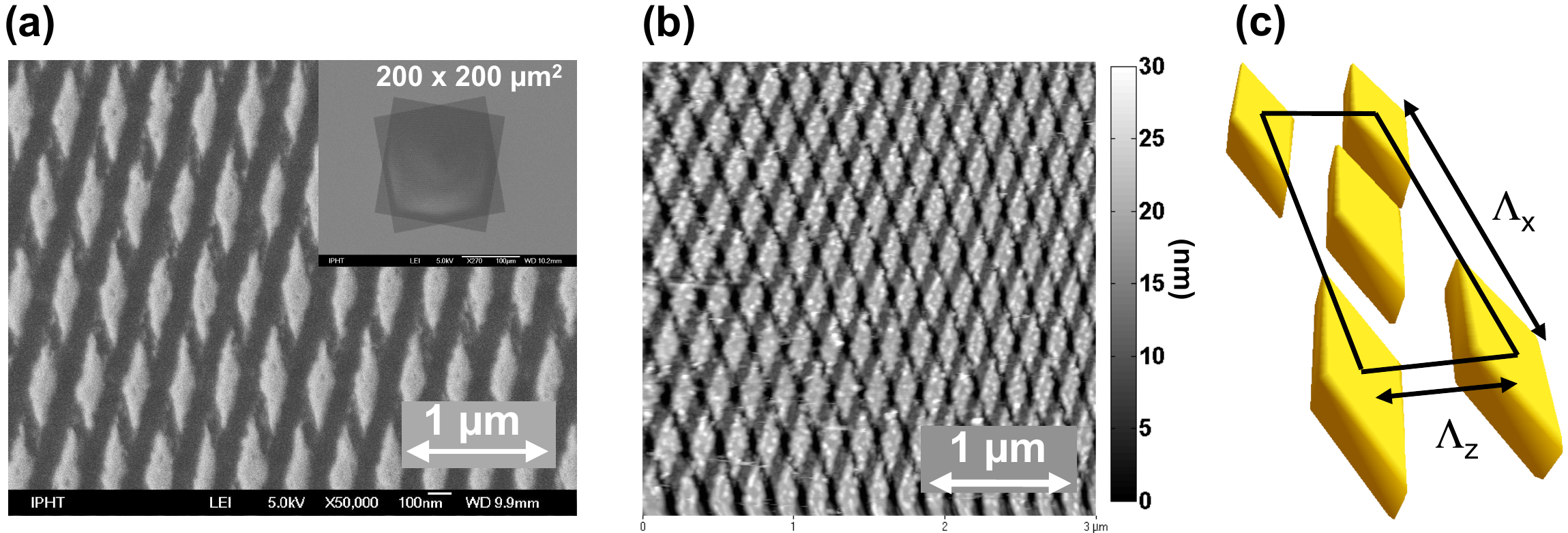}
\caption{(Color online) \textbf{(a)} A scanning electron microscopy (SEM)
image of a manufactured array of nanoantennas. The inset shows a low
resolution image where the two tilted gratings used in the
fabrication process can be identified. \textbf{(b)} The atomic force
microscopy (AFM) investigation of the same sample provides the same
rhombic-shaped material distribution as already shown in
\textbf{(a)}. Additionally, one can measure and verify the
manufactured thickness of the Au films which corresponds to the
desired $20~nm $. \textbf{(c)} The resulting unit cell of the
manufactured structure is marked by the black line.}
\label{pics_manufacturing}
\end{figure}

To fabricate nanostructures that meet these requirements we applied
electron beam lithography of suitable masks together with a dry
etching technique. The entire process is reported in \cite{Huebner}.
Here we briefly summarize the principal fabrication steps. At first,
a 20 nm gold (Au) layer was evaporated on a fused silica substrate.
The electron-beam resist material (PMMA) was then spin-coated on top. It was
afterwards exposed in a crossed-grating approach with two
one-dimensional grating masks. Subsequently, the Au layer was etched
by an ion beam (Ar). In a final step the remaining resist material
was removed by means of oxygen plasma stripping. Parameters of the final structure which are subject
to variations and which can be controlled in the fabrication process
are the duty cycle, the tilting angle of the two gratings, and their
period. The careful adjustment of all such fabrication parameters allows to control the plasmonic
properties of the samples. For illustrative purposes
Fig. (\ref{pics_manufacturing}) shows selected examples of fabricated
structures. As a result, the final structure is composed of
periodically distributed rhombi with lateral dimensions fixed by the
grating parameters used for the resist exposure. This allows for the
desired fast, reproducible, and large area manufacturing of the
optical nanoantenna arrays. Another advantage concerning the
application of such substrates as efficient SERS structures is the
high density of hot spots resulting in a defined field enhancement
across the entire sample. Besides the high density of hot spots the alignment of the
nanoantennas on a regular grid can also be used to launch PSPPs.
Their excitation introduces a further resonance into the system. PSPPs
can be excited since their momentum mismatch to free space is
compensated by the reciprocal grating vector. \\
\begin{table*}[t]
\begin{center}
\begin{tabular}{lccc}
\hline\hline Parameter   &
                                                                                                          			\textbf{Sample 1}   	&\textbf{Sample 2}  &\textbf{Sample 3}  \\ \hline
Period in \textit{x}-direction $\Lambda_x$ $(\mu \mathrm{m})$			&  0.437			&   0.800			& 1.243			\\
Period in \textit{z}-direction   $\Lambda_z$ $(\mu \mathrm{m})$		&  0.198			&   0.187			& 0.376			\\
Rhombus length $\Delta x$   $(\mu \mathrm{m})$					&  0.279			&   0.514			& 0.810			 \\
Rhombus width $\Delta z$    $(\mu m)$							&   0.085			&   0.125			& 0.247			\\
Resonance wavenumber $x$-pol. $1/\lambda$ $(\mathrm{cm}^{-1})$  	&   8,547			&   5,470			 & 5,700			\\
Resonance wavenumber $z$-pol. $1/\lambda$ $(\mathrm{cm}^{-1})$  	&   16,780			&   16,340 ; 13,890	& 15,480 ; 11,600	\\
\hline\hline
\end{tabular}
\end{center}
\caption{Detailed parameters of the three nanoantenna samples
with different manufacturing parameters that were selected for
further investigation. Sample 1 represents a regular array of
rhombs, which can be realized by a $34^{\circ}$ tilt of both
illumination gratings. Sample 2 and 3 are characterized by roughly the same apex angle
for the nanoantenna of ($27^{\circ}$ and $34^{\circ}$, respectively) but they are fabricated with an increased
period of both crossed gratings. This translates into a larger
period of the two-dimensional nanoantenna array as well as an
expansion of the nanoantenna dimensions itself. The thickness of the Au
layer is a final parameter that can be used to tailor the plasmonic
properties of the samples. In all samples it was chosen to be $20$ nm.}
\label{table_1}
\end{table*}
Taking advantage of this process various samples have been fabricated and
thoroughly characterized. In the following we restrict ourselves to
three samples. They were selected since they cover the main physical
effects occurring in our nanoantenna arrays. Spectral features of
other samples are comparable; though they may slightly differ in
their resonance wavelengths. Detailed parameters of the three
relevant samples are shown in Tab. (\ref{table_1}).

\section{Optical characterization and simulations}
\begin{figure*}[t]
\centering
\includegraphics[width=13cm,angle=-0]{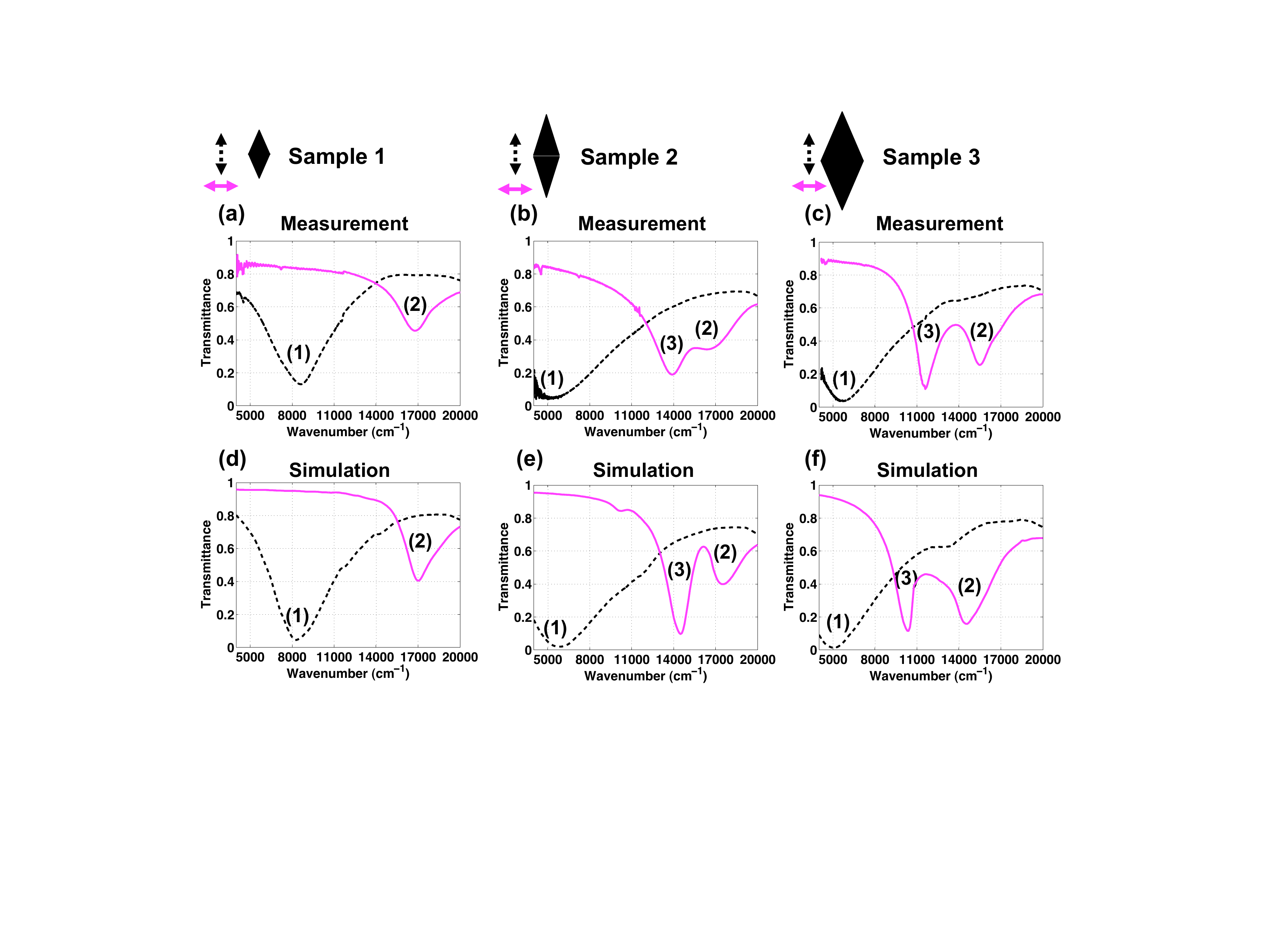}
\caption{(Color online) Measured transmission spectra for sample 1
\textbf{(a)}, sample 2 \textbf{(b)}, and sample 3 \textbf{(c)}.
Sample 1 represents a 2D arrays of gold rhomboids
sustaining two plasmonic eigenmodes. They can be excited depending
on the polarization of the incident electric field with respect to
the nanoantenna. For the black-dashed line the incident electric
field is parallel to the long axis. For the magenta-solid line it is
parallel to the short axis. The spectral position of the resonances
are indicated by (1) and (2), respectively. By increasing the size
and the period of the 2D nanoantenna array, as done in sample 2 \textbf{(b)} and 3 \textbf{(c)},
a third resonance appears. It is understood as a propagating surface
plasmon polariton. The comparison to the respective numerically
(FMM) simulated spectra is shown in \textbf{(d-f)}.} \label{pics_spectral_measurements_FMM}
\end{figure*}

Besides the structural characterization with SEM and AFM [shown in
Fig. (\ref{pics_manufacturing})], we characterized all samples by
conventional optical far-field transmission measurements
\cite{PerkinElmer} at normal incidence. Results are shown in
Fig. (\ref{pics_spectral_measurements_FMM}).
In the measured transmission spectra of all three samples
resonances can be
identified which depend on the polarization of the incident electric
field [Fig. (\ref{pics_spectral_measurements_FMM}a-c)]. 

For the rhomb-like nanoantenna sample 1 only the two
plasmonic dipole resonances associated to the principal axis of the
structure can be observed (labelled as 1,2). They appear due to the different
transverse dimensions at different wavenumbers around
$8,547~ \mathrm{cm}^{-1}$ and $16,780~ \mathrm{cm}^{-1}$.
Additionally, sample 1 exhibits a very sharp apex tip along its
major axis which will lead to an increased near field. Consequently,
sample 1 represents already a favorable sample that could be used as
an efficient SERS substrate. By increasing the period of the
illumination gratings, the period of the unit cell as well as the
rhombi dimensions increase simultaneously. This was done for sample
2 and 3 of which the spectral properties are shown in
Fig. (\ref{pics_spectral_measurements_FMM}b,c).
The spectral response
is slightly modified when compared to sample 1. At first,
we observe two resonances that occur at $5,470~ \mathrm{cm}^{-1}$ (1)
and $16,340~ \mathrm{cm}^{-1}$ (2) for the two polarizations. Similar
to sample 1, they are determined by the lengths of the principal
axes of the rhombi. However, in addition, a third resonance occurs (for sample 2 at
$13,890~ \mathrm{cm}^{-1}$ and for sample 3 at $11,600~ \mathrm{cm}^{-1}$) for an incident
polarization of the electric field parallel to the short axis.
For an even larger rhombus, as represented by sample 3, resonance (3) can be observed more pronounced [Fig. (\ref{pics_spectral_measurements_FMM}c)], since it is now clearly separated from resonance (2).
\\
In order to investigate the origin of this additional resonance (3) and to
verify our measured data we performed simulations of the spectral
response of all samples. These simulations were done by using the
Fourier Modal Method \cite{FMM}. This method solves Maxwell's equations without approximation by taking explicit advantage of the periodic nature of the sample. In these three-dimensional simulations the dispersive material properties as well as the exact particle shape including the substrate and ambient media are fully taken into account. Results are show in
Fig. (\ref{pics_spectral_measurements_FMM}d-f). The spatial
dimensions of the nanoantennas as well as information about period,
metal thickness, etc. have been extracted from the SEM and AFM
images of the fabricated samples. They were fully taken into account
in the simulations. Material data for the Au dielectric function
was taken from literature \cite{JC}. By comparing numerical and
experimental spectra we observe a good agreement with
respect to the resonance positions and the resonance widths for all
excited modes. Particularly, all resonances for sample 2 and 3 are also
revealed in the simulations.\\
Prior to applying the samples for SERS measurements, we want to prove
the plasmonic origin of all occurring resonances. Together with the
numerical description, we apply two analytical methods to interpret
our observed resonances. We show that due to apparent similarities
the resonances denoted as (1) and (2) can be similarly treated as
localized dipole modes of the two main axes of the rhombus, while
resonance (3) can be interpreted as a grating excited PSPP. For this
purpose we describe the scattering response of the individual
rhombus within the electric dipole limit in an effective medium approach \cite{BH}. It allows to predict the appearance
of the LSPP modes. On the contrary, we motivate the mode denoted as
(3) as a PSPP mode that is launched due to the alignment of the
nanoantennas in a grid. We do so by investigating its angle and
period dependent excitation frequency \cite{PSPPandLSPP}. The
dispersion relation which we will reveal for this mode corresponds
to that of a PSPP at a metal-dielectric interface. For a precise
analytical prediction of this dispersion relation that is compared
to numerical results, one only has to take care that the metallic
structure is a composite material whose properties can be reliably
derived on the grounds of an effective medium theory. Details are given below.\\
\section{Analytical considerations (LSPP)}
\begin{figure}
\centering
\includegraphics[width=8.7cm,angle=-0]{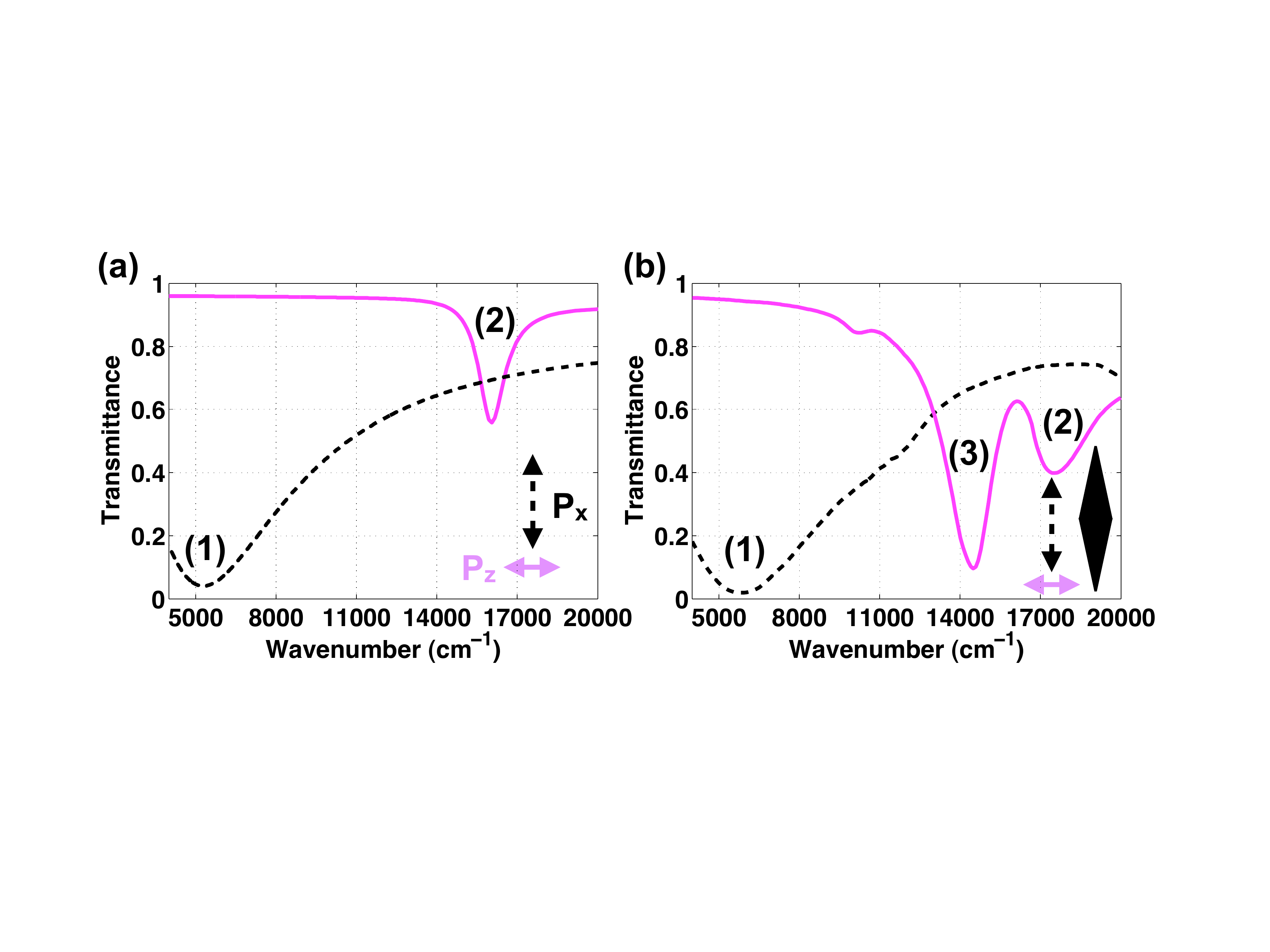}
\caption{(Color online) The quasi-statically \textbf{(a)} and the
numerically \textbf{(b)} obtained transmission spectra for an
incident electric field polarized along the long (axis 1) and the
short axis (axis 2) of sample 2 are presented. It can be seen that
resonance (1) and (2) as the two main axes modes are excellently
reproduced in both simulations. Additionally, resonance (3) is only
present in the numerical calculations.} \label{pics_QS}
\end{figure}

At first we want to describe the LSPP modes in detail. For this
purpose, we employ an analytical approach that is compared to the
rigorous simulations. From the above comparison of sample 2 and sample 3
we might anticipate that the two resonances (1) and (2)
are associated with LSPPs. However, the verification that the
resonances can be predicted by relying on analytical grounds
constitutes an inevitable tool in the design of SERS
substrates with predefined resonance wavelengths.\\
To treat the problem analytically, we assume that we can describe the plasmonic properties of an isolated rhombus by its dipole moments. The structure is assumed to be biaxial anisotropic; reflecting the different length of the axes.
Assuming that this approximation is valid we can calculate the scattering
response of an ensemble of individual rhombic nanoantennas. For this purpose we follow an approach known from the fields of metamaterials in terms of an associated effective medium, comprising electric dipole interaction. We mention that an electric dipole resonance results in a strongly dispersive effective permittivity with a Lorentzian lineshape as
\begin{align}
\label{Eq1}
&\epsilon_{\text{eff}}(\omega)_l=\epsilon_\infty+\frac{A_l}{\omega_{0l}^2-\omega^2-i\gamma_l\omega}.
\end{align}
In Eq. (\ref{Eq1}) $\omega_0$ is the resonance frequency, $A$ is the phenomenological  resonance or oscillator strength, $\gamma$ the damping and $\epsilon_\infty$ the background permittivity for infinite frequencies (here vacuum; $\epsilon_\infty=1$). The introduced damping constant accounts for both radiative and non-radiative losses and therefore ensures the correct resonance width. Within the dipole limit we have the possibility to account for the fundamental resonances that are characterized by a unique resonance frequency $\omega_{0l}$ ($l\in [x,z]$) in conjunction with the resonance width and strength. All these parameters are phenomenologically introduced. In the next step we adjust the free parameters of Eq. (\ref{Eq1}) in order to reproduce the measured spectral properties of the samples. For this purpose transmission through a slab of material with a permittivity as given in Eq. (\ref{Eq1}) is calculated and matched to the measured spectra by adjusting the free parameters in a least square sense.  For obtaining the transmittance spectra shown in Fig. (\ref{pics_QS}) the following parameters have been applied. $A_x=40.5\cdot 10^{30}s^{-2},A_z=1.9\cdot 10^{30}s^{-2},\omega_{0x}=9.8018\cdot 10^{14}s^{-1},\omega_{0z}=3.0159\cdot 10^{15}s^{-1},\gamma_{x}=2.8274\cdot 10^{14}s^{-1},\gamma_{x}=1.6965\cdot 10^{14}s^{-1}$.
In Fig. (\ref{pics_QS}) the transmission spectra of such an effective medium including a possible set of parameters are presented. It can be seen that by means of this approximation only two resonances can be obtained, while a third resonance is missing. In passing we note that the third dipolar resonance of the third axis naturally cannot be probed at normal incidence.\\
In the next step we will show that this empirically introduced permittivity covering the localized dipole modes of the rhombi is sufficient to derive the origin of the third resonance, which is an effect of the alignment of the nanoantennas in a regular grid, mimicking a PSPP resonance.

\section{Analytical considerations (PSPP)}
\begin{figure*}
\centering
\includegraphics[width=13cm,angle=-0]{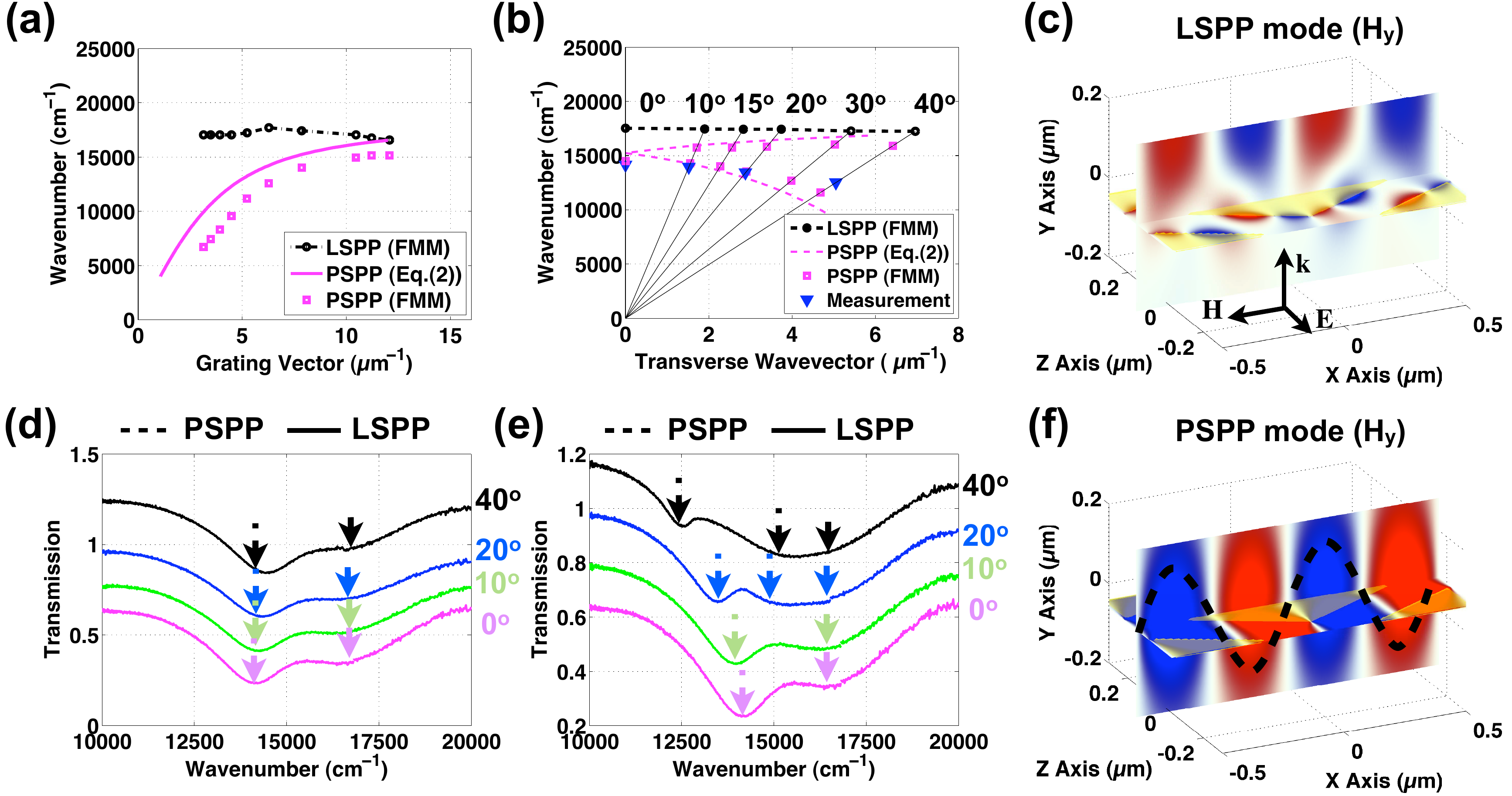}
\caption{(Color online) \textbf{(a)} Variation of the resonance position
as a function of the inverse lattice period in $z$ direction: magenta squares - FMM, magenta solid lines -  quasi-PSPP [Eq. (\ref{Eq2})].\textbf{(b)}
Wavenumber splitting of resonance (3) due to the variation of the angle of incidence $\theta$. magenta squares - FMM, magenta solid lines -  quasi-PSPP [Eq. (\ref{Eq2})], blue triangles - measured resonance positions. The black dashed lines show the effect of lattice period and angle variation on the LSPP resonance (2) which obviously remains invariant.\textbf{(d)} The measured angular dependence of the transmittance for a tilting angle along the $x$ axis (long rhombus axis) and the $z$ axis (short rhombus axis) \textbf{(e)}. The LSPP resonance position remains unchanged weather the quasi-PSPP resonance is sensitive to an additional $k$ vector in $z$ direction. The transmittance has been added with $0.2$ for each incidence angle, in order to preserve clarity. Finally snapshots of time harmonic sequences, calculated with FDTD for the near field distribution of sample 3 for the LSPP \textbf{(c)} (Media 1) and the PSPP frequency \textbf{(f)} (Media 2) are presented ($H_y$ component). A standing wave pattern for the PSPP resonance emerges and is underlined by the black-dashed line.} \label{pics_PSPP}
\end{figure*}
In contrast to the analysis before, which was performed with the
intention to elucidate the character of the LSPP eigenmodes, we
proceed now by investigating the properties of the quasi-PSPP
eigenmodes in detail. For this purpose we consider the nanoantenna
array as an effective medium with the specified optical properties
expressed by the previously introduced effective electric permittivity $\epsilon_\text{eff}(\omega)$ [Eq. (\ref{Eq1})]. To
observe PSPPs on such an effective medium interface the transverse
momentum and the energy conservation must be satisfied. These
conditions can be cast into the equation \cite{Ebbesson_fomula}
\begin{align}
&\frac{\omega}{c}\sqrt{\frac{\epsilon_{\text{eff}_l}(\omega)\epsilon_\text{d}(\omega)}{\epsilon_{\text{eff}_l}(\omega)+\epsilon_\text{d}(\omega)}}
=\frac{\omega}{c}\sin(\theta)+m_j\frac{2\pi}{\Lambda_j}.
\label{Eq2}
\end{align}
The left hand side of Eq. (\ref{Eq2}) corresponds to the transverse
PSPP wavevector \cite{Raether} at an interface between an effective
medium described by the effective permittivity fixed by the polarization of the incident light
$\epsilon_{\text{eff}_l}(\omega)$ and a dielectric medium with a
permittivity $\epsilon_\text{d}(\omega)$. In contrast to an ordinary
PSPP at a metal dielectric interface the permittivity of the metal
was replaced by the previously derived effective permittivity. The corresponding mode
is known as a spoof plasmon \cite{VidalScience} or quasi-PSPP mode.
The first term on the right hand side accounts for the transverse
wavevector of the incident illumination where $\theta$ is the
angle of incidence. In addition to the transverse momentum provided
at oblique illumination there is another transverse wavevector
introduced by the 2D grating (second term on the right-hand side).
Taking into account the grating vectors in Eq. (\ref{Eq2})
does not contradict the effective medium approach. The
combination of both effects, i.e. a non-vanishing grating vector and
the effective medium treatment of the nanoantenna arrays is covered
by the \textit{empty lattice approximation}
\cite{empty_lattice}. Thus the integer $m_j$ denotes the
diffraction order, while $\Lambda_j$ is
the respective period of the unit cell as shown in
Fig. (\ref{pics_manufacturing}c).

To test the quasi-PSPP character of the observed resonance (3) we
performed numerical calculations (Fourier Modal Method - FMM) as well as transmission measurements for angular incidence which were compared with
predictions from Eq. (\ref{Eq2}). It was done by computing
numerically or measuring the frequency of excited resonances (taken as the dip in
transmission) depending on a certain parameter (chosen period or
angle of incidence). These resonance frequencies are compared to
predictions from Eq. (\ref{Eq2}).\\
At first the grating period for normal incidence ($\theta=0^\circ$)
was varied numerically. This causes the grating vector of the lattice to be the
only contribution to the transverse wavevector in
Eq. (\ref{Eq2}). The equation provides the resonance frequencies for
each grating vector, see Fig. (\ref{pics_PSPP}a).
Here we found theoretically and experimentally that the resonance position for the quasi-PSPP mode is only dependent on the magnitude of the period pointing in $x$ direction Fig. (\ref{pics_PSPP}d,e), while the effective permittivity is fixed due to the polarization of the incident beam to $\epsilon_\text{eff}(\omega)_z$.
Consequently,
we set $m_x=1$ and $m_z=0$. For the ambient permittivity
$\epsilon_\text{d}(\omega)$ in Eq. (\ref{Eq2}) we selected air ($n=1$).
Comparing the expected dispersive behavior according to Eq. (\ref{Eq2}) with the numerical one we observe some deviations in terms of a constant offset between both, but a qualitative agreement in the overall shape  Fig. (\ref{pics_PSPP}a).
We point out that the rhombus dimensions were fixed for the grating vector variation, thus the choice of the lattice period can be applied to tailor the resonance position for the quasi-PSPP mode in a wide spectral interval. In turn this variation of the unit cell period increases or decreases the number of rhombi in a predefined volume. This will mainly result in a change of the oscillator strength $A$ in Eq. (\ref{Eq1}), since this value accounts for the resonance weighting of the individual rhombus embedded in an effective medium. Considering this effect the predictions for the resonance frequency positions by a fixed effective permittivity can be considered as an approximation to motivate the precise calculated resonance dependence.
\\
In a second step the incidence angle was varied, which modifies the
first term on the right hand side of Eq. (\ref{Eq2}). The dispersion
relation of the quasi-PSPP mode used for the grating period
variation was applied without any further adaption. We observed
a better agreement between the quasi-PSPP description and the
numerically performed analysis [Fig. (\ref{pics_PSPP}b)] as for the grating vector variation since the effective material parameters now remain fixed and only the excitation conditions were varied. In order to ensure the PSPP origin of the observed resonance we also performed measurements of the sample for a variable angle of incidence Fig. (\ref{pics_PSPP}d,e) to see the numerically predicted resonance behavior.
The coincidence between the measured and the analytically as well as numerically calculated resonance positions verifies the assumption that the observed mode is not a localized mode, since for such modes the resonance positions are invariant under varied excitation conditions. Such variations only influence the excitation strength of the respective LSPP mode. Here we observe a resonance dependence that can be completely explained with a PSPP excitation and significantly differs from the behavior of a localized mode.
Finally we performed FDTD simulations with a spatial resolution of $2.5~nm$ and the same material parameters as for the FMM simulations in order to calculate the near field distributions for both modes, rigorously. Therefore we selected sample 3 since here both modes are spectrally separate, which permits the undisturbed observation of the LSPP and the PSPP field patterns Fig. (\ref{pics_PSPP}c,f) (Media 1, 2). It can be clearly seen that the LSPP near fields are determined by localized features at the rhombus surface, while we observe a standing wave pattern in the perpendicular ($x$) direction with respect to the incident polarization direction ($z$) for the PSPP excitation condition. We mention that we plotted the field component which is normal to the surface ($H_y$) and thus is not contained within the excitation field ($E_z,H_x$). Additionally we emphasize that the propagation direction differs from the propagation direction observed for PSPP modes excited with one-dimensional gratings. There the PSPP mode propagates parallel to the incident polarization. Due to the two-dimensional nanoantenna grating reported here the PSPP mimicking mode is sensitive to modifications of the grating vector perpendicular to the polarization direction as observed in the resonance dependence Fig. (\ref{pics_PSPP}b). \\
Based on the dispersive behavior observed in the
far field simulations and the observed near field distributions, we identify resonance (3) as a surface plasmon
polariton excited by the grating. We refer to it as a quasi-PSPP
since the metallic surface is not flat and the metal itself constitutes an effective medium. In all calculations, the resonances (2) and (1) (not shown here) remain unaffected by any
variation of the period or the illumination conditions of the
arrays. Thus they are plasmonic
features of the individual
nanoantenna itself.\\
\section{Application to SERS}
Finally, we employ the most promising nanoantenna array (sample 2)
for SERS measurements at optical frequencies. It is most promising
as it is doubly resonant in the spectral region of interest around
$15,000~\mathrm{cm}^{-1}$ ($660~\text{nm}$). We emphasize that by doubly resonant we understand here the presence of both; the PSPP as well as the LSPP within a predefined spectral interval. Stipulated by the fact
that both elaborated plasmonic resonances are associated with
strongly enhanced local fields, we expect an improvement of the SERS
signal for this particular nanoantenna array. The main advantage of
the sample is the possibility to have a strong field enhancement at
the excitation wavelength as well as at the emission wavelength by
incorporating two different resonances. 
this is the first investigation on SERS signals with structures
sustaining such double plasmonic resonances.
Moreover, the spectral
separation between both resonances can easily be modified by
varying, e.g., the angle of incidence or the period of the lattice.
This introduces a further degree of freedom to tune the spectral
position of the resonance to match
to a particular SERS application.\\
\begin{figure*}
\centering
\includegraphics[width=13cm,angle=-0]{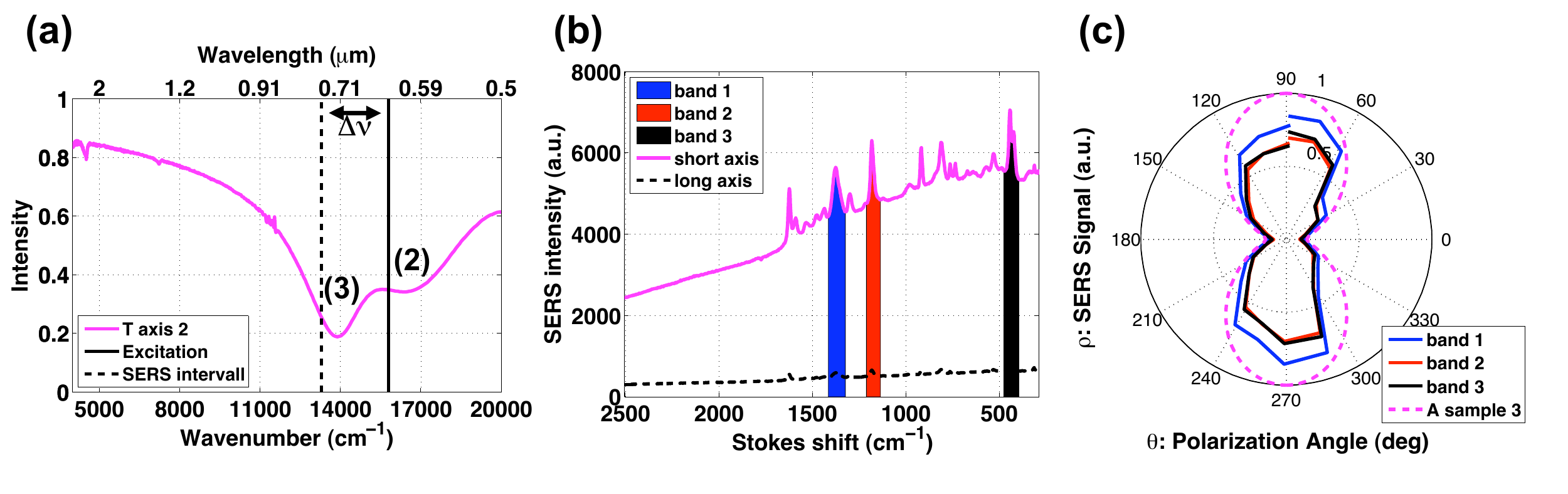}
\caption{(Color online) \textbf{(a)} The SERS excitation wavenumber (black
solid line) and the SERS measurement interval $\Delta\nu$ (black
dashed lines) together with the measured short axis resonances of
sample 2 are shown. Both resonances (2) and (3) are included in the
SERS measurement interval. \textbf{(b)} The SERS intensity for a
polarization along the short axis and the long axis of sample 2 is
presented. Furthermore, three different bands for the angular SERS
measurements are indicated. \textbf{(c)} The polarization angle
resolved SERS measurement is shown for the three aforementioned
bands.
} \label{pics_sers_measurements}
\end{figure*}

For the SERS measurements the triphenylmethane dye crystal violet as a
typical SERS analyte was selected.
Fig. (\ref{pics_sers_measurements}a) shows the excitation
wavenumber (black solid line) which was realized by a $\text{HeNe}$
laser ($15,798~\mathrm{cm}^{-1}$) as well as the wavenumber window
for the SERS measurement (black dashed lines). One can see that the
spectral domain where SERS measurements were done are nicely covered
by both plasmonic resonances. The excitation wavelength is close to
the LSPP resonance, while the SERS interval (Stokes shift) includes
the quasi-PSPP resonance. The resulting SERS signal for our
first measurement is shown in Fig. (\ref{pics_sers_measurements}b).
In two different experiments the polarization of the incident laser
radiation for the excitation was aligned along either of the two
principal axis of the nanoantenna. The extracted enhancement factors
for several well selected bands were found to be in the order of
$10^3$. We note that there are several possibilities to define the
enhancement factor. We have applied the definition for the averaged
\textit{SERS surface enhancement factor} (SSEF) given in
\cite{SERSenhancementfactor}.
\begin{align}
\text{SSEF}(\omega) =
\frac{I_{\text{SERS}}(\omega)c_{\text{RS}}H_{\text{eff}}}{I_{\text{RS}}(\omega)\mu_\text{M}\mu_\text{S}A_\text{M}}
\label{eq_SERS_EF}
\end{align}
In Eq. (\ref{eq_SERS_EF}) $I_{\text{SERS}}(\omega)$ is the SERS
intensity and $I_{\text{RS}}(\omega)$ is the Raman signal of crystal
violet without the nanonatenna array. The two parameters
$\mu_\text{S}$ and $\mu_\text{M}$ are the densities of the molecules
on the nanoantenna array and of the nanostructures per unit area,
respectively. $H_{\text{eff}}$ is the effective height of the
scattering volume and $c_{\text{RS}}$ corresponds to the
concentration for the reference Raman signal measurement. One of the
most sensitive factors in this equation is the effective SERS area
$A_\text{M}$. We have used an effective area of $0.02\times 0.02~\mu
\mathrm{m}^2$ which corresponds to the mean curvature area of the
small axes edges of the rhombus in the SEM images. To complement our
investigations and to support the assumptions on this effective
area, we have also performed finite-difference time-domain
simulations to investigate the near fields in detail. From these simulations we estimated an effective enhanced near field area of equal size as well as an electric near field increase of $|E/E_0|^4\approx 10^3$ as we have experimentally observed for the SERS enhancement factor. We mention that these numerical estimations represent an approximation, since the exact shape and the corresponding field enhancement factor are strongly dependent on the computational discretization parameters, e.g. the mesh size and the corner radii. Nevertheless the obtained field enhancement factor correlates with the experimentally observed SERS enhancement factor. \\
In addition to the SERS measurement with a polarization along the
short rhombus axis we have rotated the polarization of the
illumination systematically in order to investigate the polarization
dependence of the SERS signal. As it was shown recently
\cite{SERSangular}, such polarization resolved SERS measurements can
be used to elucidate the plasmonic origin of the field enhancing
elements, which are represented by the nanoantenna arrays in our
samples. Results of our investigations are summarized in
Fig. (\ref{pics_sers_measurements}c). The observed angular dependence
of the measured SERS signal corresponds to the numerically
calculated absorption dependence at the SERS excitation wavenumber.
We note that the absorption is maximal at our samples when the
transmission tends to be minimal. Thus, absorption peaks as well as
transmission dips indicate plasmonic resonances in our system. We
observe a strong correlation between the excitation of plasmonic
resonances and an enhanced SERS signal for all three selected SERS
bands. This can be seen from Fig. (\ref{pics_sers_measurements}c).
The spectral positions of the bands are indicated in
Fig. (\ref{pics_sers_measurements}b). Moreover, the angular
dependence has approximately a $\sin^2(\theta)$ distribution. This
is expected for our dipole-type nanoantennas, since the excitation
strength of a dipole is proportional to the sine of the polarization
angle of the exciting electric field normal to the dipole,
translating into a $\sin^2(\theta)$ dependence for the intensity.
The fact that the SERS enhancement factor is larger than unity even
for the polarization direction along the long axis
($\theta=0^{\circ}$) is attributed to a non-vanishing field
enhancement due to the nanoantennas sharp edges for this
polarization even for the long rhombus axis.\\
\section{Conclusion}
In summary, we have presented a plasmonic nanoantenna array that
exhibits localized and propagating plasmon polariton modes. A
comprehensive theoretical investigation incorporating numerical and
analytical calculation as well as experimental verifications has been presented to elucidate the nature
of the observed modes. We have shown that an excellent agreement
with measured transmission spectra is obtained. Finally, we have applied the
nanoantenna samples for SERS measurements in the optical domain in
order to exploit the specific advantages of the double resonant
structures. We have observed an enhanced SERS signal exactly for the
polarization directions where resonances are excited in the
nanoantennas. In addition to the polarized SERS measurement along
the principal axes of our structures we have presented SERS
experiments with continuously rotated polarization direction of the
illuminating laser radiation. The resulting polarization dependent
SERS signal reproduces the corresponding absorption dependence,
indicating that the SERS signal strongly correlates with the
excitation of the plasmonic eigenmodes. This work was performed with
intention to introduce a new multimode plasmonic nanoantenna system
allowing highly controllable and large-area plasmonic substrates
with tailored resonance positions suitable for SERS measurements.
Therefore, Au was chosen as a highly biological compatible metal
even if Ag is a better plasmonic
material in the optical domain regarding damping and field enhancement.\\
\section*{Acknowledgements}
Financial support by the Federal Ministry of Education and Research
(Innoregio-ZIK, InnoProfile-JBCI and Metamat) as well as from the state of Thuringia
in the ProExcellence programm is acknowledged.

\end{document}